# Acceleration of solar wind ions to 1 MeV by electromagnetic structures upstream of the Earth's bow shock

K. Stasiewicz[1,2,6], S. Markidis[3], B. Eliasson[4], M. Strumik[2] and M. Yamauchi[5]

[1] *Swedish Institute of Space Physics, Uppsala, Sweden*
[2] *Space Research Centre, Polish Academy of Sciences, Poland*
[3] *Royal Institute of Technology, PDC Center for High Performance Computing, Stockholm, Sweden*
[4] *Institute of Theoretical Physics, Ruhr-University Bochum, Germany*
[5] *Swedish Institute of Space Physics,Kiruna, Sweden*
[6] *Physics and Astronomy, University of Zielona Gora, Poland*

**Abstract** – We present measurements from the ESA/NASA Cluster mission that show in situ acceleration of ions to energies of 1 MeV outside the bow shock. The observed heating can be associated with the presence of electromagnetic structures with strong spatial gradients of the electric field that lead to ion gyro-phase breaking and to the onset of chaos in ion trajectories. It results in rapid, stochastic acceleration of ions in the direction perpendicular to the ambient magnetic field. The electric potential of the structures can be compared to a field of moguls on a ski slope, capable of accelerating and ejecting the fast running skiers out of piste. This mechanism may represent the universal mechanism for perpendicular acceleration and heating of ions in the magnetosphere, the solar corona and in astrophysical plasmas. This is also a basic mechanism that can limit steepening of nonlinear electromagnetic structures at shocks and foreshocks in collisionless plasmas.

**Introduction.** – It is generally believed that the acceleration of particles to super thermal energies in the inner heliosphere is done mainly by shock waves [1, 2]. These may be driven by solar flares and coronal mass ejections, and are also formed at planetary bow shocks, and at boundaries between streams of fast and slow solar wind plasma in co-rotating interaction regions [3, 4]. The origin of the energetic ions observed upstream of the Earth's bow shock [5] is generally attributed to shock drift acceleration [6] of solar ions or to leakage from the magnetosphere [7]. In this letter, we present measurements that show in situ acceleration of ions to energies of 1 MeV outside the bow shock, incompatible with both the concepts of leakage and shock drift acceleration. The observed heating can be associated with the presence of electromagnetic structures with gradients of the electric field that lead to ion gyro-phase breaking and to rapid, stochastic acceleration of ions in the direction perpendicular to the ambient magnetic field. This heating mechanism prefers heavy ions, which is supported by the measurements reported in this paper. Our results may be directly applicable to the currently disputed problem of ion heating in the solar corona [8,9] and ion energization in magnetospheric plasmas [10].

**Observations.** – The Cluster mission [11] provides measurements from the Earth's bow shock, foreshock, and solar wind regions out to the apogee at 19 $R_\mathrm{E}$ (Earth radii) by four identical spacecrafts. A comprehensive set of instruments makes it possible to study particle acceleration processes at the bow shock using the unique, multi-point capabilities of the mission. In Fig. 1 we present an overview of Cluster-1 measurements during crossings of the magnetosheath, bow shock, foreshock and solar wind regions on 27-28 January 2003. Figure 1a shows the speed of ions measured by the CIS-HIA experiment [12] and Fig. 1b the total magnetic field measured by the FGM [13] experiment. The position of the apogee at a distance of 19 $R_\mathrm{E}$ [16, 10, -2] GSE is reached at 17:04 UT and marked with a vertical line in Fig. 1a, while crossings of the bow shock occur at positions marked with vertical lines labeled 'B1' and 'B2' in Fig. 1c. The records show classical textbook signatures of the solar wind being slowed down by the bow shock from 450-500 km/s to 200-300 km/s and an associated magnetic field compression from about 5 nT in the solar wind to 30 nT in the magnetosheath – a layer behind the bow shock. The measured temperature of solar wind ions (not shown) is increased by a factor of 20 from 0.1 MK in the solar wind to 2 MK in the magnetosheath. Figure 1c shows fluxes of energetic ions: $H^+$ protons, $He^+$, and heavy ions from the $(C, N, O)^+$ group. The heavy ions are present in the 948-1414 keV instrument channel of the RAPID experiment [14]. It is seen that the region of appearance of energetic ions is detached from the bow shock and it is located far out in the solar wind.

The regions near the crossing of the bow shock do not have measurable fluxes of energetic ions at energies >100 keV, in contradiction to the theories that bow shocks are major suppliers of super thermal particle populations in the solar system. Neither does the absence of measurable fluxes of energetic ions in the magnetosheath support hypothesis of a leakage from the magnetosphere. Recent simulations [15] also show that quasi-perpendicular shocks are basically incapable of generating nonthermal particles from the thermal solar wind plasma, while quasi-parallel shocks can increase the ion energy 10-fold, from ∼1 to ∼10 keV, still far below the observations of ∼MeV ions.

Figure 1d shows the angle between the magnetic field vector and the radius vector of the satellite position  (a proxy to the shock normal direction on the day side). It can be seen in Fig. 1d that bow shock crossings have quasi-perpendicular orientation of the magnetic field, but the ion heating occurs in regions with quasi-parallel orientation of **B**. The Cluster orbit shown in Fig. 1 without appreciable fluxes of energetic particles at the bow shock crossings but with detached regions of hundreds keV ions in the solar wind represents a rather typical picture of Cluster observations as can be seen already in the Quick Look Plots (e.g. http://www.cluster.rl.ac.uk/).

We notice in Fig. 1b that the presence of high energy ions heated to ∼ 1 MeV is associated with regions of low values of the magnetic field with fluctuations, where the amplitude of $B$ oscillates between 3-10 nT. A detailed inspection of the heating regions shows that fluxes of accelerated ions (Fig. 2a) are associated with compressive magnetic structures, where the enhancements of the magnetic field are in phase with compressions of the plasma density (Fig. 2b), and with fluctuations of the electric field (Fig. 2c) in the form of steepened structures. The $E_y$ component of the electric field [16] in Geocentric Solar Ecliptic (GSE) coordinates is shown here for all four spacecrafts separated by 3000-5500 km.

Magnetic pulsations described above represent typical foreshock phenomena,  called also SLAMS [17], and have been associated with ion-ion right-hand resonant instability [18,19]. According to this model, the instability arises from interactions between the solar wind beam and a hot population of ions heated at the bow shock and streaming sunward, away from the bow shock. However, in the heating regions we observe large ion pressure anisotropy, which may also produce the measured electromagnetic structures via fire-hose instability [20]. The observed structures can represent fast alfvenon solutions for quasi-parallel propagation of nonlinear waves in Hall-MHD approximation to anisotropic plasma [21, 22].

In Fig. 3 we show ion velocity distribution function measured in the region where most energetic ions were observed. It shows an intense solar wind beam with a velocity of about 500 km/s and a diffuse heated ion population. It appears that the energization of ions to higher energies $\sim 10 - 1000$ keV must occur outside the bow shock, in the regions of electromagnetic structures shown in Figs. 1 and 2. In Figure 4 we show differential particle flux as a function of energy measured by the HIA and CODIF detectors, which are parts of the CIS instrument, and high energy ions measured by the RAPID experiment. It is seen that heavier ions possess higher energy fluxes than ions $H^+$ and $He^+$ indicating that the heating mechanism favours ions with larger masses. The energetic heavy ions observed by the RAPID instrument using solid-state detector are well beyond the one-count level, and furthermore they are also observed in the CIS data using time-of-flight detector.

The electron measurements made on Cluster (not shown here) exhibit a rather constant electron temperature ($\sim$0.15 MK) throughout the ion heating events, which excludes current-driven instabilities and parallel electric fields that would preferentially heat electrons.

**A universal ion acceleration mechanism.** – In the absence of broadband turbulence in the ion heating region we assume a hypothesis that the acceleration mechanism is related to gradients of the electric field which can lead to fast ion heating perpendicular to the magnetic field, as observed in laboratory plasma [23] and in the aurora [24, 25]. An electric field gradient in the direction of propagation of a structure may lead to the breaking of the gyro-phase motion of the ion and to the onset of chaos, i.e. trajectories of two closely spaced particles start to diverge exponentially and they acquire large transverse velocities. A necessary condition for this type of acceleration of particles with charge $q$ and mass $m$ is [23, 25, 26]

$$\left|\frac{\partial E_y}{\partial y}\right| \geq \frac{q}{m} B_0^2, \quad (1)$$

where the differentiation is in the direction of propagation $\hat{\mathbf{y}}$ of the solitary structures. The above condition is most easily met by single charged heavy ions in a weak magnetic field. It implies the following relations

$$\frac{1}{\omega_{ci} B}\frac{\partial E_y}{\partial y} \sim \frac{1}{\omega_{ci}}\frac{\partial V_x}{\partial y} \sim \frac{V_x}{\omega_{ci} L_y} \sim \frac{R_d}{L_y} \geq 1, \quad (2)$$

which means that the bulk ion flow has velocity shear with gradient scale $L_y$ that is smaller than the direct ion gyroradius defined as $R_d = V_x/\omega_{ci}$, with $\omega_{ci}$ being the ion cyclotron frequency. The electric field gradient required by Eq. (1) is $2.4 \times 10^{-9}$ Vm$^{-2}$ for protons and $1.5 \times 10^{-10}$ Vm$^{-2}$ for oxygen ions. Such gradients would correspond to a variation of the electric field by 0.24 (0.015) mV/m over the distance of 100 km for $H^+$ ($O^+$), respectively. Gradients of this magnitude can be found in Cluster measurements made in the heating regions, providing experimental support for the acceleration mechanism discussed in this paper. We would also like to note that steepening of electromagnetic structures seen in Fig. 2 could eventually produce arbitrarily strong gradients that would fulfil equation (1). Thus, the ion heating mechanism discussed in this paper could represent the basic dissipation mechanism controlling the nonlinear steepening process in collisionless plasmas.

Ion heating to high energies (>100 keV) is observed in solar wind regions with small magnetic fields, but not at the bow shocks where $B$ is increased, in agreement with conditions (1-2). Observations that minority ions from the (C, N, O)$^+$ group acquire higher fluxes and energies than the more abundant $H^+$ population corroborates also this conclusion.

**Test particle simulations.** – To demonstrate that the heating mechanism on two-dimensional electric field gradients is indeed operational we perform test particle simulations using the equation of motion for a charged particle with velocity $\mathbf{v}$, mass $m$, charge $q$

$$m\frac{d\mathbf{v}}{dt} = q(\mathbf{E} + \mathbf{v} \times \mathbf{B}), \quad \frac{d\mathbf{r}}{dt} = \mathbf{v}, \quad (3)$$

where $\mathbf{B} = B_{z0}\hat{\mathbf{z}}$ is the external magnetic field and $\mathbf{E} = -\nabla\Phi$ the electric field obtained from the potential of the form

$$\Phi(x,y) = L_y E_{y0} \cosh^{-2}(x/L_x) \cosh^{-2}(y/L_y) + xE_{x0}. \qquad (4)$$

Here $L_x$ and $L_y$ represent the spatial scales of the field gradients. The constant electric and magnetic fields $E_{x0}$ and $B_{z0}$ move the particles in the y-direction into a lattice of electric field structures with strongest gradients in the $E_y$ GSE components of the electric field as seen in Fig. 2c.

Figure 5 shows sample trajectories for two ions in case of two-dimensional structures. It is seen that strong gradients break the gyromotion and the ions increase their perpendicular velocities. The ion heating is evident from the increase of the ion gyro-radius after each crossing of the electric field structure. The maximum energy gain occurs for the particle starting from the right and it is 1500 times the initial kinetic energy. When the condition (1) is not fulfilled the ion perpendicular energy does not change after encounters with the electric field structures. Similar heating is also observed in case of one-dimensional structures obtained in the limit $L_x \to \infty$ in equation (4). The electric field model represents potential wells that can trap colder ions from the main population and prevent them from being accelerated by the mechanism discussed above, while only faster ions can enjoy free acceleration. The simulation model with a constant magnetic field is an idealisation of observations which show variations of both amplitude and the direction of the magnetic field. The simulations are done in the reference frame of the nonlinear structures, which move with respect to the plasma with a super-alfvenic velocity.

The energy source for the acceleration process is the electric field and the electric potential related to the bulk flow of the solar wind. An average electric field in the ion heating regions is $\approx 1.6$ mV/m (RMS) that would produce an accelerating potential of 10 kV over a distance of $1R_E$, which correspond to a typical spatial size of a single nonlinear structure that can be regarded as a fast alfvenon [22]. The energy of solar wind protons at a speed of 500 km/s is only 1.3 keV. Thus, a spatial displacement over a distance of 10 $R_E$ is required to accelerate a portion of solar wind ions to 100 keV. Considering the fact that regions covered with alfvenons are very large and last many hours in Cluster data, it is possible that some ions will meander in fluctuations of the electric field over a distance large enough to acquire the energy of 1 MeV seen in the measurements. In fact, the nonlinear structures are sweeping space while carried by the solar wind with some relative speed. An equivalent distance of 100 $R_E$ required to accelerate ions to 1 MeV corresponds to 30 minutes exposure time for super-thermal ions immersed in the flow of nonlinear electromagnetic structures at a relative speed of 350 km/s. Without performing extensive simulations we cannot determine whether the heating of ions to 1 MeV in the foreshock requires pre-heating to $\sim 10$ keV at the bow shock and subsequent escape to the parallel foreshock, or the heating process can be accomplished locally on the tail of the solar wind population only.

**Conclusions.** – In summary, we have shown that steepening process of nonlinear electromagnetic structures in the Earth's foreshock region may lead to fulfilment of the condition (1) for the gradient of the electric field that would cause ion gyro-phase breaking and rapid, stochastic acceleration of ions in the direction perpendicular to the ambient magnetic field. Meandring of ions across the electromagnetic moguls created in the terrestrial foreshock region can lead to increase of their kinetic energy to 1 MeV over a cumulative distance of $\sim 100\, R_E$. The ion gyro-phase breaking and acceleration mechanism illustrated in Fig. 5 represents a basic universal dissipation and damping mechanisms for electromagnetic structures formed in collisionless plasmas. This mechanism may also be responsible for perpendicular acceleration and heating of ions in the solar corona, interplanetary, and astrophysical plasmas.

∗ ∗ ∗

This work has been partly supported by the Polish National Science Centre (DEC-2012/05/B/ST9/03916). B. E. acknowledges support from the Deutsche Forschungsgemeinschaft (DFG) through Project SH21/3-1 of Research Unit 1048.

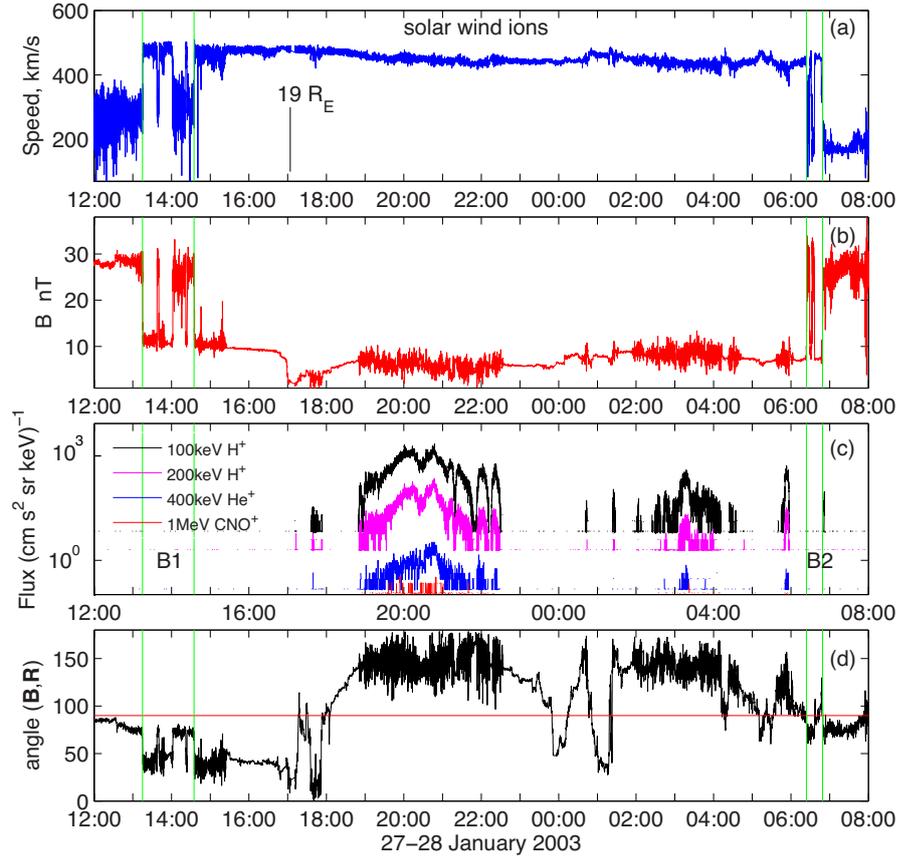

Fig. 1: An overview of Cluster-1 (C1) crossings of the magnetospheric boundaries during 27-28 January 2003. Panels (a) and (b) show the ion speed and magnetic field strength, respectively. They exhibit typical behaviour at quasi-perpendicular bow shocks, where the solar wind streaming at 450-500 km/s is slowed down to 200-300 km/s in the magnetosheath, while the magnetic field is compressed from 5 nT in the solar wind/foreshock region up to 30 nT in the magnetosheath. Panel (c) shows fluxes of energetic ions in the energy range 92-1414 keV, which occur in regions distinctively different from the bow shock crossings marked with vertical lines labeled B1 and B2. Panel (d) shows angle between the radial direction (a proxy to the shock normal direction) and the magnetic field vector.
.

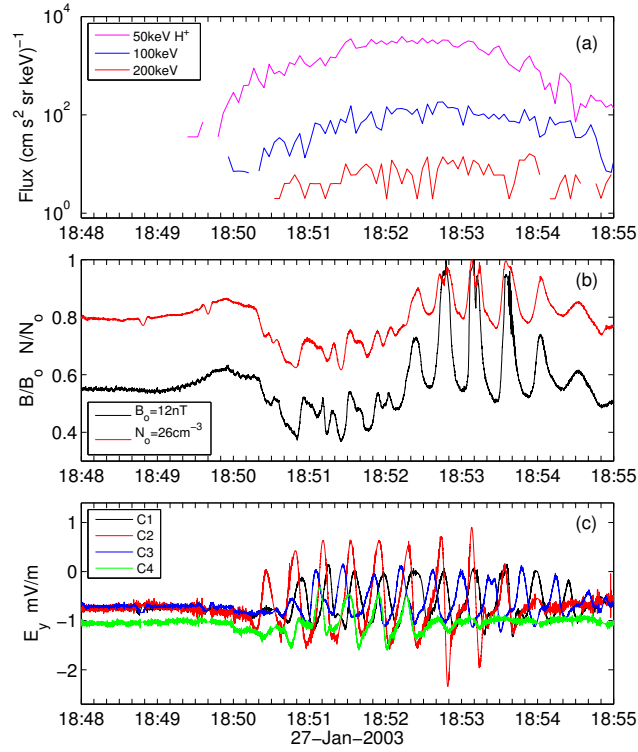

Fig. 2: Details of the data from Fig. 1. Panel (a) shows fluxes of energetic protons observed by C1 in association with trains of non-linear structures seen in the measurements of the magnetic field and plasma density seen in panel (b), and $E_y$ GSE component of the electric field shown in panel (c) for all four spacecrafts. Sufficiently steep gradients of the electric field would break ion gyro-motion and lead to the onset of chaos and rapid, stochastic heating of ions.

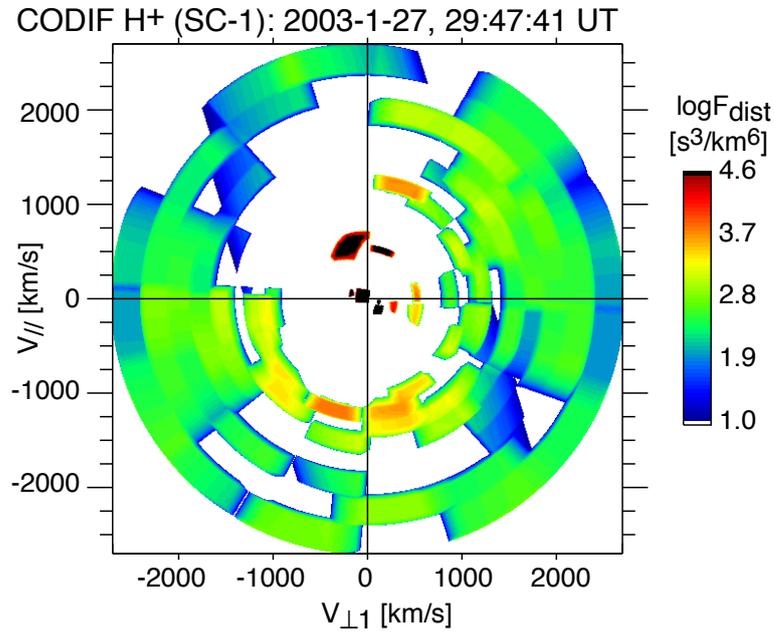

Fig. 3: Ion velocity distribution function measured by the CODIF instrument on Cluster-1 at 20:47 UT, where most energetic ions were measured (see Fig. 1c). The picture shows an intense solar wind proton beam with a velocity about 500 km/s surrounded by a diffuse energised ion population.

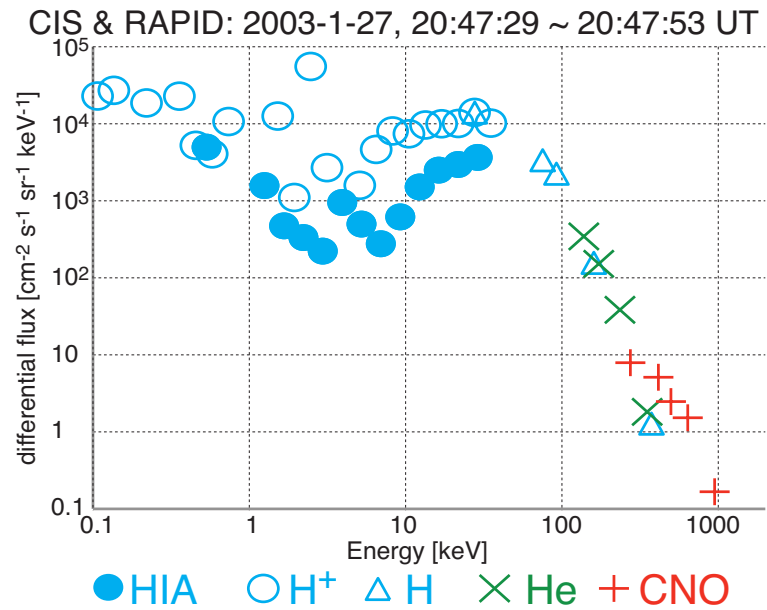

Fig. 4: Differential ion flux measured at 20:47 UT. The plot shows overlap of data taken from, respectively, three ion detectors: HIA ($H^+$), CODIF ($H^+$), and RAPID [$H^+$, $He^+$, and heavy ions from the $(C, N, O)^+$ group] on the Cluster-1 spacecraft. It is seen that heavy ions have higher fluxes and energies than ions $H^+$ and $He^+$.

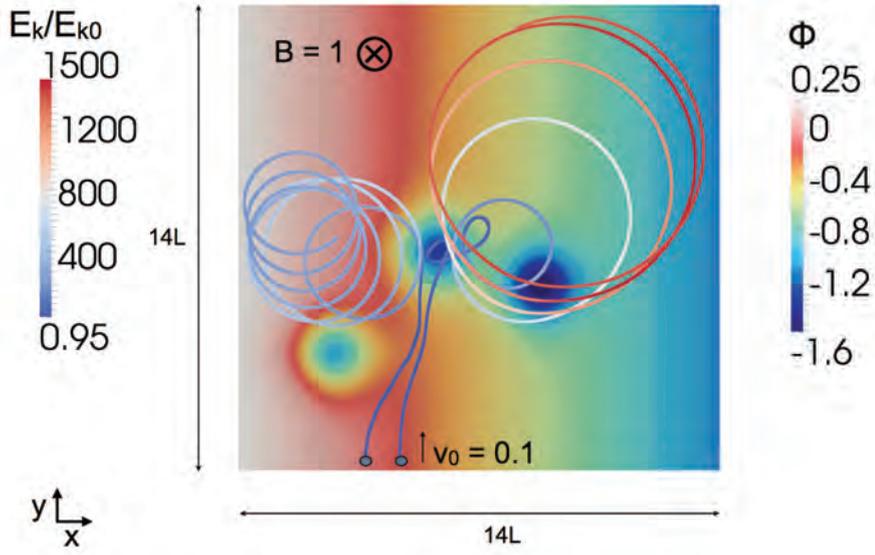

Fig. 5: Two ion trajectories superimposed on $\Phi$ contour-plot that represents two dimensional field of moguls (Eq. 3 with $L_x = L_y$). Time is normalised with the ion gyro frequency $\omega_i$ and length with $L$. The initial ion velocity is 0.1 $L\omega_i$ and the amplitude parameter implied by (1) is $(E_{y0}L^{-1})(mq^{-1}B_{z0}^{-2}) = 1$. The ion heating via a two dimensional field of moguls is evident from the increase of the ion gyro-radius after each encounter with a mogul. The colour of the particle trajectory is the kinetic energy normalised to the initial particle kinetic energy (0.01 the amplitude the potential well).